\documentclass{article}

\usepackage[table]{xcolor}
\usepackage[utf8]{inputenc}
\usepackage{tismir}
\usepackage{amsmath}
\usepackage{hyperref}
\usepackage{url}
\usepackage{graphicx}
\usepackage{booktabs}
\usepackage{lipsum}
\usepackage{pifont}
\newcommand{\cmark}{\ding{51}}%
\newcommand{\xmark}{\ding{55}}%
\newcommand{\omark}{\ding{109}}%

\title{From Imitation to Innovation: The Divergent Paths of Techno in Germany and the USA}
\author{%
Tim Ziemer\thanks{Institute of Systematic Musicology, University of Hamburg, Hamburg, Germany},%
~and Simon Linke\thanks{Ligeti Center, Hamburg University of Applied Sciences, Hamburg, Germany}}

\date{}

\authorref{Ziemer,~T. and  Linke,~S.}
\authorshort{Ziemer and Linke} 

\begin{document}


\twocolumn[{%
\maketitleblock
\begin{abstract}
Many documentaries on early house and techno music exist. Here, protagonists from the scenes describe key elements and events that affected the evolution of the music. In the research community, there is consensus that such descriptions have to be examined critically. Yet, there have not been attempts to validate such statements on the basis of audio analyses. In this study, over 9,000 early house and techno tracks from Germany and the United States of America are analyzed using recording studio features, machine learning and inferential statistics. Three observations can be made: 1.) German and US house/techno music are distinct, 
2.) US styles are much more alike, and 3.) scarcely evolved over time compared to German house/techno regarding the recording studio features. These findings are in agreement with documented statements and thus provide an audio-based perspective on why techno became a mass phenomenon in Germany but remained a fringe phenomenon in the USA. Observations like these can help the music industry estimate whether new trends will experience a breakthrough or disappear.
\end{abstract}
}]
\saythanks{}


\section{Introduction}
In 2024, techno culture in Berlin became an UNESCO Intangible Cultural Heritage\footnote{See \href{https://www.unesco.de/staette/technokultur-in-berlin/}{https://shorturl.at/T28nq}.} in recognition of over $30$ years of a vivid and still active scene. In contrast, the techno exhibition at the Michigan State University Museum\footnote{See \href{https://museum.msu.edu/msu-museum-presents-techno-the-rise-of-detroits-machine-music/}{https://shorturl.at/Az6yz}} is looking back at the important role Detroit used to play in 20th century techno. Techno flourished in Germany but remained an underground scene in the United States \citep{planetwissen}.

In 1994, techno became mainstream music in Germany with over $25,000$ visitors at the Mayday rave in April 1994 in Dortmund \citep[p. 81]{szene}, and over $100,000$ people joining the Loveparade techno event in Berlin in 1994 (cf. \citep[p. 116]{szene}).
. 

In the field of Music Information Retrieval (MIR), big data audio analysis of Electronic Dance Music (EDM) has provided insights into the music. But very few studies connect these findings with music history, scenes, and judgments by protagonists. Based on interviews with protagonists, musicological studies have revealed differences that may explain the breakthrough in Germany and the lack of a breakthrough in America. But musicological studies lack a critical review of protagonists' statements through audio analysis. The motivation of this study is to bridge this gap between MIR and musicological studies. In the study, we stick to the sound terminology of \citep{Ziemer2024}, i.e., we distinguish between audio engineering and psychoacoustical terminology.

\subsection{Previous Work}

\subsubsection{Music Information Retrieval}
Many studies analyzed the audio files of EDM. \citep{yadati} used rhythm features and machine learning to find the drop in dance music, \citep{hockman} trained classifiers to determine which breakbeat has been sampled in drum and bass tracks. \citep{collins} examined influence of funk, disco, synth-pop, electro/hip hop, punk/post-punk on 1980s Detroit techno and Chicago house using predictive models. They analyzed $248$ tracks. Through supervised classifiers and unsupervised clustering of timbre features, they found that both styles were much alike and shared a disco and synth pop heritage.

\citep{caparrini} classified subgenre of dance music tracks using, e.g., random forest classifier, trained on $92$ audio features. They achieved an accuracy of up to $59$\% for $10$ genres, and their boxplots showed that beats per minute (bpm) was the most relevant feature. The authors speculate that ``house'' is an ill-defined subgenre, as the classifiers have a low recall for house.

\citep{popli} classified Electronic Dance Music (EDM) subgenres using spotify's audio metadata from $15,000$ songs equally distributed over $5$ subgenres (house, drum and bass, techno, hardstyle and trap). Multiple Machine Learning (ML) classifiers, like random forests, achieved an accuracy between $0.833$ and $0.913$ Their motivation was to automate the process of homogeneous playlist generation. 

\citep{tanyarin} extracted recording studio features from $1,841$ EDM tracks. Using a random forest classifier, they could predict which out of $10$ famous disk jockeys would play which track with an accuracy of $0.63$, underlining that sound plays an important role in EDM, and that recording studio features represent relevant aspects of the sound.

\citep{knees} evaluated tempo estimation algorithms on $664$ dance music tracks. They found that the tempo was correct in no more than $77$\% of all cases. The authors lament that there is still a lack of EDM-specific annotated data sets.

\citep{mauch} raised the issue that much of what has been written about popular music is anecdotal rather than scientific. They analyzed the US Billboard Hot $100$ between 1960 and 2010 using audio features and machine learning to study the history of popular music using big data audio analysis. Inspecting a self-similarity plot, they identified three revolutions: years where the preceding years were very different, but following years were very similar. They identified one of these revolutions as the rise of British Rock music in America in 1964.

Overall, studies from the field of Music Information Retrieval (MIR) provided valuable insights into house and techno music, but these insights are rarely put in relation with the development of the music scene.

\subsubsection{Musicology}
Musicological studies on house and techno music exist.

\citep{lothwesen} analyzed two techno tracks to highlight how commercialization of techno went along with rudimentary sound design, arrangement and rhythm.

\citep{wilderom} aimed at answering the question of why dance music became so popular in the UK but not the USA. They argue, e.g., that the Rolling Stone Magazine in America was conservative, while the British NME magazine was trend-seeking. As a consequence, an underground phenomenon was promoted to the public through a music magazine in the UK but not in the US.


Many documentaries and (auto-)biographies reflect on the development of house and techno music in Germany and the United States. Some publications go further and mix personal experience with interviews and researched facts, like \citep{flash,familie,volkweintechno}.

Some protagonists' statement on German and US house and techno music can be found repeatedly, summarized in Table \ref{tab:observations}.

\begin{table*}[tb]
    \centering
    \begin{tabular}{|c|c|c|c|}\hline
    Statement & MANOVA & SOM & RF\\\hline
    Early German house emulated American house. \citep{volkweintechno} & \omark & \cmark & \omark \\
    Germans found their techno in 1992. \citep{technodeep}& \cmark & \cmark & \omark \\
    The German scene was more dynamic. \citep{loveparade,sicko} &  \omark & \cmark & \cmark \\
    German megaraves made all music sound the same. \citep{wick}&  \omark & \xmark & \xmark \\
    America failed at establishing extraordinary styles. \citep{flash}& \omark & \cmark & \cmark \\
    Detroit's second generation techno sounded European. \citep{flash} & \omark & \cmark& \omark\\
    The fall of the Berlin wall was a catalyst for techno's breakthrough. \citep{technodeep}& \omark& \omark&\omark\\\hline
    \end{tabular}
  \caption{Protagonists' statements on German and US house and techno music. Check-marks anticipate which statement will be supported by our audio analyses (cf. Sect. \ref{verification}). MANOVA, SOM and RF are our analysis methods described in Section \ref{method}.}
  \label{tab:observations}
\end{table*}

While appreciated as a source of information, researchers agree that documentaries, biographies, and interviews with protagonists have to be reflected critically \citep{volkweintechno,lothwesen} 
and that it is important to put music analysis in relation to its social space \citep{hawkins} 
and the scene \citep{szene}
. In addition, \citep[p. 5]{volkweintechno} points out the need to search for agreement between statements of protagonists and music analyses.

Overall, musicological studies on house and techno music provided insights into the scenes and their developments and relations. But the studies tend to evaluate interviews and biographic material without putting it in relation with the audio content. Only a few tracks are analyzed, if any.

Our approach is to combine MIR and musicology approaches: Big data music analysis as commonly dome in MIR studies, put in relation with the scene, as demanded by musicologists. 




\section{Method}
\label{method}
We analyze American and German house and techno music based on recording studio features. Then, we relate the results with protagonists' statements. Instead of testing each statement with a dedicated method (which is inefficient and prone to the multiple comparisons problem, i.e., false positives), we apply inferential, parametric statistics (Multivariate Analysis of Variance, MANOVA), nonlinear data exploration (Self Organizing Map, SOM) and a nonlinear classifier (Random Forest, RF). These methods provide an overall picture, complement each other and allow for qualitative observation (visual inspection) and quantitative audio-based analyses. The features and the methods are described below.

\subsection{Material}
The HOTGAME \citep{hotcorpus} 
corpus is analyzed. It contains over $9,000$ tracks from CDs and records of the first author and his affiliation. The distribution is fairly even ($4667$ German and $4362$ US tracks). It was collected based on the artist and record label names that can be found in the literature (e.g.\citep{Anz1995,sicko,volkweintechno,technodeep}). Only music derived from the house music branch was considered, including (deep/garage/acid/hip) house, techno, (goa) trance, ambient, eurodance, (happy) hardcore, schranz, gabber, drum \& bass and tekkno, a particularly hard style of techno in Germany. 
The data was labeled by the first author, based on booklets, the cited literature, and supplemented by web research (including \url{discogs.com}). The data was randomly checked by his students. All tracks have been released between 1984 and 1994 and come from Germany or the US. The assignment was based on the nation where the producer grew up. Several metrics provide evidence that 
the HOTGAME corpus is representative for house and techno music from 1984 to 1994 \citep{hotgame}. For example, the collection contains music from most dance music labels of that time ($97$ German and $85$ US labels), and the number of tracks correlates with the number of active record labels in the respective country.


The HOTGAME corpus does not include the audio files, but recording studio features that quantify aspects of the sound that are also monitored in the recording studio when producing such music. These recording studio features have proven their value in quantifying the loudness war \citep{mainstream}, telling different music genres apart \citep{daga}, characterizing the unique sound profile of hip hop producers \citep{nikita}, and explaining which electronic music DJ would play which track \citep{tanyarin}.

Recording studio features from the HOTGAME corpus are analyzed using Multivariate Analysis of Variance (MANOVA), a conventional method in musicology research (cf. \citep{manovamusic}), and machine learning, a typical method in music information retrieval (cf. \citep{peeters}). The music from both nations is analyzed in terms of differences between the nations, the temporal development over the years, and in terms of various house and techno styles.

The analysis results are reported in the \emph{Results} section. In the \emph{Discussion}, they are put in relation to statements from protagonists of the scenes, aiming at validating narratives that describe the German and American house and techno scenes. Furthermore, we reflect on the benefit of applying methods from the field of music information retrieval for answering questions from the field of musicology.

\subsection{Features}
There is little consensus about techno analysis methods \citep{gfpm}. However, musicians are considered to produce tracks rather than compose music. This means that sound, sound design and audio effects play the essential role, while traditional compositional aspects such as instrumentation, lyrics, 
melody, harmony 
and rhythm 
may play a minor role (see e.g. \cite{Hemming2016,produktion,hawkins}). This focus is derived from the DJ practice of music mixing and enhancement with samples and audio effects. The music is born in the recording studio, not in the rehearsal room on a piano or in a jam session.
%

Consequently, audio analysis tools from the recording studio are utilized, as they represent relevant sound aspects. In this study, four audio analysis tools are utilized\footnote{The feature extraction code is available on \url{https://timziemer.github.io/technoanalysis.html}.}: 
\begin{enumerate}
    \item bpm
    \item PhaseSpace
    \item ChannelCorrelation
    \item CrestFactor
\end{enumerate}

The tempo in terms of beats per minute (\emph{bpm}) is chosen by music producers and is an important characteristic of many styles \citep{honingh}. 
The \emph{PhaseSpace} quantifies the point cloud of a phase scope. Phase scopes indicate the volume of a track in terms of the sound pressure level, panning, and the stereo distribution \citep{Stirnat2017}. It is distilled to a single value at each time frame using box counting. The \emph{ChannelCorrelation} is often included in phase scope tools. Originally utilized to monitor mono compatibility, music producers use it to estimate the stereo width of a mix \citep{current}. The \emph{CrestFactor} is the ratio of peak and root-mean-square level. Percussive sounds tend to have a much larger crest factor than complex tones, drums have a larger crest factor than atmospheric sounds, and dynamic range compression reduces the crest factor \citep{mainstream}. In this study, each track is represented by a feature vector that holds each feature's median value. These median values are normalized corpus wide such that each feature has a mean value of $0$ and a standard deviation of $1$. 

The recording studio features are not arbitrary. Music producers set the bpm in the sequencer, so it represents a conscious decision. The other features, like the channel correlation and the root-mean-square pressure, are not set. But firstly, they result from production methods (like the choice of synthesizers and audio effects, and their settings). Secondly, they are usually inspected during the music production process using audio analysis tools \citep{tanyarin}. While the first is true for all audio features (including timbre features like MFCCs and zero-crossing-rate as well as rhythmic or melodic features), the second does not hold for many features (like MFCCs and zero-crossing-rate). Thirdly, the recording studio feature magnitudes are directly interpretable. For example, a low channel correlation implies a high stereo width. Fourthly, these features exhibit no significant correlations with each other, which is an important prerequisite for MANOVA and Self Organizing Maps. These three strengths make the recording studio features meaningful for early house and techno music, which is producer-driven and focuses strongly on sound and less on harmonic progression or other conventional compositional aspects.

For each track, the median values of the recording studio features are stored and analyzed.

\subsection{Inferential Statistics}
Inferential statistics, like MANOVA, are frequently applied in musicology studies (cf. \citep{manovamusic}). 
MANOVA tests whether a number of dependent variables (hence \emph{Multivariate}) differs between groups. In the case of a two-way MANOVA, the groups are formed by two independent variables, in our case nation and year. Additionally, difference between their combination (nation*year) is analyzed. The dependent variables are the recording studio features. MANOVA quantifies which features are significantly different between groups, applying the F-test. The null hypothesis of a MANOVA is that the distributions of feature magnitudes are equal between nations and/or years. MANOVA quantifies differences by a significance value and effect size, but it does not reveal much about the nature of the difference. Effectively, MANOVA is similar to applying multiple ANOVAs, but assures a higher statistical power. However, as some assumotions of MANOVA do not hold in our dataset (e.g., the distributions are not normal and exhibit heteroscedasticity), it is wise to review the results critically, using complementary methods (cf. \citep{mcshane}), like boxplots, self-organizing maps and random forest explained in the following.


\subsubsection{Boxplots}
Boxplots summarize each feature's distribution over nation and year to enhance the interpretability of the MANOVA through visual inspection. They indicate median, lower and upper quartile and outliers in terms of items with $1.5$ times the interquartile range.

However, as MANOVA and boxplots focus on distributions, they do not reveal any information about individual tracks. Self-organizing maps present information on every single track while representing the topology of the distributions.

\subsection{Self-Organizing Map}
Aggregating features and representing individual tracks is the strength of Self-Organizing Maps (SOMs). SOMs characterize each individual track by its feature magnitudes. This enables the inspection of individual tracks and a nonlinear (dis-)similarity mapping. SOMs are a type of neural network \citep{kohonen}. They receive a number of high-dimensional feature vectors that represent various items, such as house and techno tracks. The output is a (usually) two-dimensional map made of units (aka. \emph{neurons} or \emph{nodes}). Every unit holds one vector that has as many dimensions as the feature vectors. On the map, every item is placed on the Best Matching Unit (BMU), i.e., the neuron on the map whose vector points most proximate to the item's feature vector. The map is trained in such a way, that the topological structure of the map resembles the topology of the original, high-dimensional feature space. In other words, the SOM puts similar items to the same, or proximate units, and dissimilar items to distant units. But it also has a zoom characteristic: if many items are very similar, they may receive a larger region, and if a single item is dissimilar, it gets only separated by a few units whose vectors point to an empty location between the item and the rest. In short, SOMs serve as a nonlinear dimensionality reduction method that loses less information than linear dimensionality reductions methods, such as projections or principal component analyses.

The SOM is usually visualized in two ways. On the \emph{Unit matrix} (\emph{U-matrix}), every unit receives a color. The color indicates the cumulated distance between the unit's vector, and all neighboring units' vectors. When they point to the same location in the high-dimensional feature space, the color is dark blue. When they point at very different locations, the color is yellow. Otherwise, the color is turquoise. So, if neighboring units are blue, they are more alike than neighboring units that are turquoise or even yellow. The \emph{component planes} represent the magnitude of every single feature over the units on the SOM. Here, the same color means that the units are similar, regarding this particular feature.

SOMs are utilized as tools for music corpus exploration \citep{Blass2019,hiphopsom,nikita,tismir,somson}. Here, they serve, e.g., as clustering tools, or user interfaces. They group music pieces that are similar and segregate music pieces that are dissimilar, according to the considered features. Moreover, SOMs provide an insight into the music and the nature of their similarity and dissimilarity.

In this study, a Self-Organizing Map is trained with all tracks from the HOTGAME corpus. The resulting map itself, the distribution of tracks by nation, by year, and by style are examined. Just like MANOVA and the boxplots, the SOM examination and additional statistics are supposed to highlight similarities and differences between tracks from Germany and America in general, over the years, and even regarding different styles.



SOMs are helpful, as they provide insight into the distribution of tracks for visual inspection. However, additional methods are required to quantify observations.
\subsection{Random Forest Classifier}
Random forest is a supervised learning algorithm. The classifier uses decision trees and Gini impurity to assign each feature vector to a class. While each decision tree tends to predict the right class for a subset of items, the collective of $100$ decision trees is able to classify most training items correctly. In this study, a random forest classifier quantifies the qualitative impression of the SOM concerning the different styles of German and American house and techno music.

\section{Results}
\label{results}
\subsection{Inferential Statistics}
Two-way MANOVA reveals significant differences between the nations ($F(4,9008) = 303$, Wilk's $\lambda = 0.8814$, $p<0.00001$, partial $\eta^2 = 0.12$), years ($F(4,9008) = 571$, Wilk's $\lambda = 0.7977$, $p<0.00001$, partial $\eta^2 = 0.20$), and nation*year ($F(21,9005) = 304$, Wilk's $\lambda = 0.8811$, $p<0.00001$, partial $\eta^2 = 0.12$) with a medium to large effect size. As a post-hoc test, Bonferroni-corrected ANOVA reveals that each of the four features differs significantly between nation, year, and nation*year ($F\geq 37$, $p<0.00001$), except the crest factor for nation*year ($F=3$, $p=0.09$). 
As MANOVA reveals significant differences, boxplots allow for qualitative insights into the distributions and the nature of their differences.

\subsubsection{Boxplots}
\begin{figure}[hbt!]
\centering
\includegraphics[width=.49\textwidth]{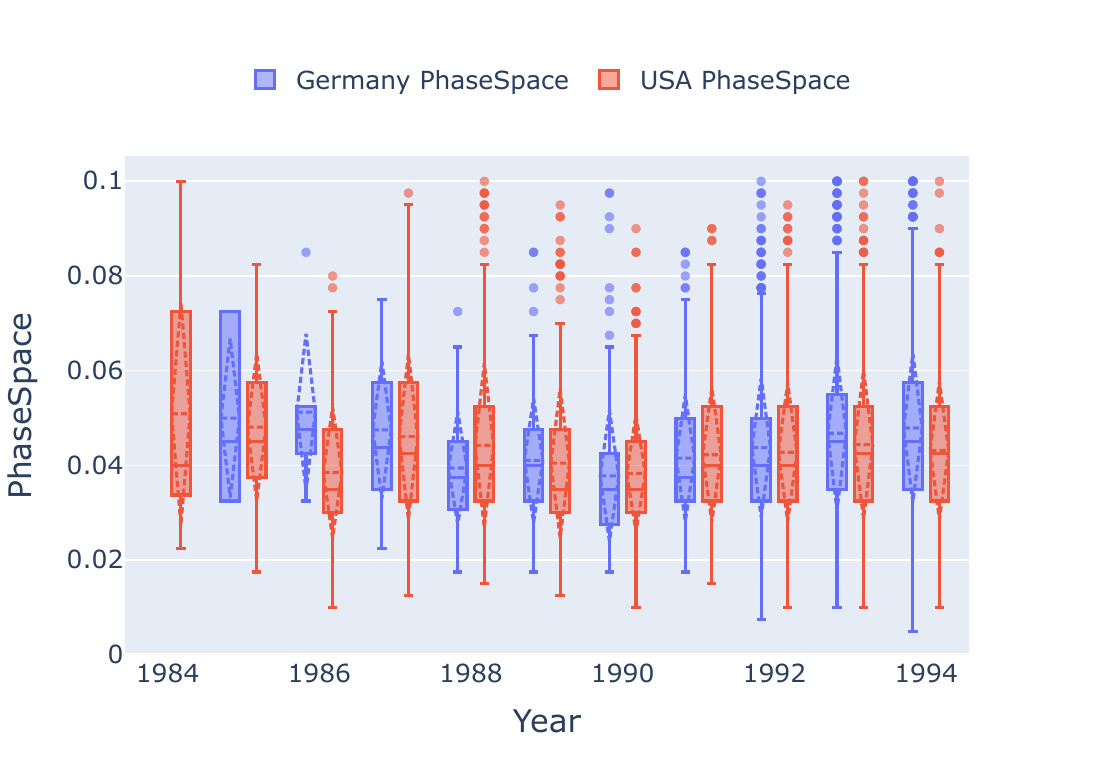}
\caption{Boxplots of median phase space scores in Germany and America over the years.}
\label{pic:boxphsp}
\end{figure}

Figure \ref{pic:boxphsp} shows boxplots of the phase space scores. On average, the phase space of German and American tracks is similar. The main difference is that there is a rise of the median value, and the whiskers tend to grow in Germany from 1990 to 1994, while both remain fairly constant in America from 1991 to 1994. This highlights that Germany had a larger and growing variety concerning dynamics and stereo mixing in that period.

\begin{figure}[hbt!]
\centering
\includegraphics[width=.49\textwidth]{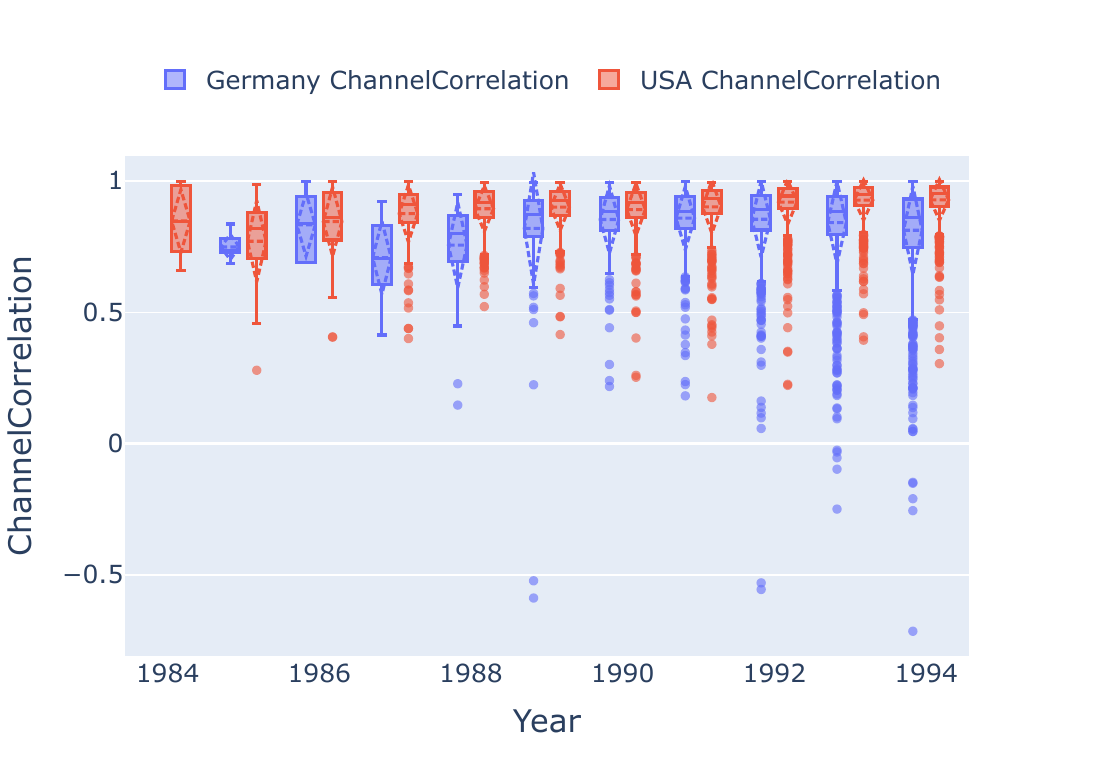}
\caption{Boxplots of median channel correlations in Germany and America over the years.}
\label{pic:boxchcorr}
\end{figure}

Box plots of the median channel correlations are illustrated in Fig. \ref{pic:boxchcorr}. In Germany, the channel correlation tends to rise until it saturates in 1990. That means the stereo width becomes narrower, probably due to a bassier sound. From 1990 on, it stays near $85$\%, but the spread increases dramatically. This means that some tracks exhibit an even higher channel correlation, while others have a much lower channel correlation. This is certainly owed to the fact that tekkno and hardcore concentrated on a heavy bassdrum (mixing in mono, i.e., high channel correlation), while trance and Eurodance added more melodic, harmonic, and atmospheric sounds (mixed with low channel correlation). 

In America, the overall channel correlation is $5$\% higher, and rises continuously. This speaks for an emphasis on mono-compatibility, which is important for nightclubs, where hard panning and other stereo techniques do not work due to the wide distribution of loudspeakers \citep[p. 268]{buch}. The spread is much smaller than in Germany, and reduces continuously. This means that the stereo width is getting more alike instead of diversifying.

\begin{figure}[hbt!]
\centering
\includegraphics[width=.49\textwidth]{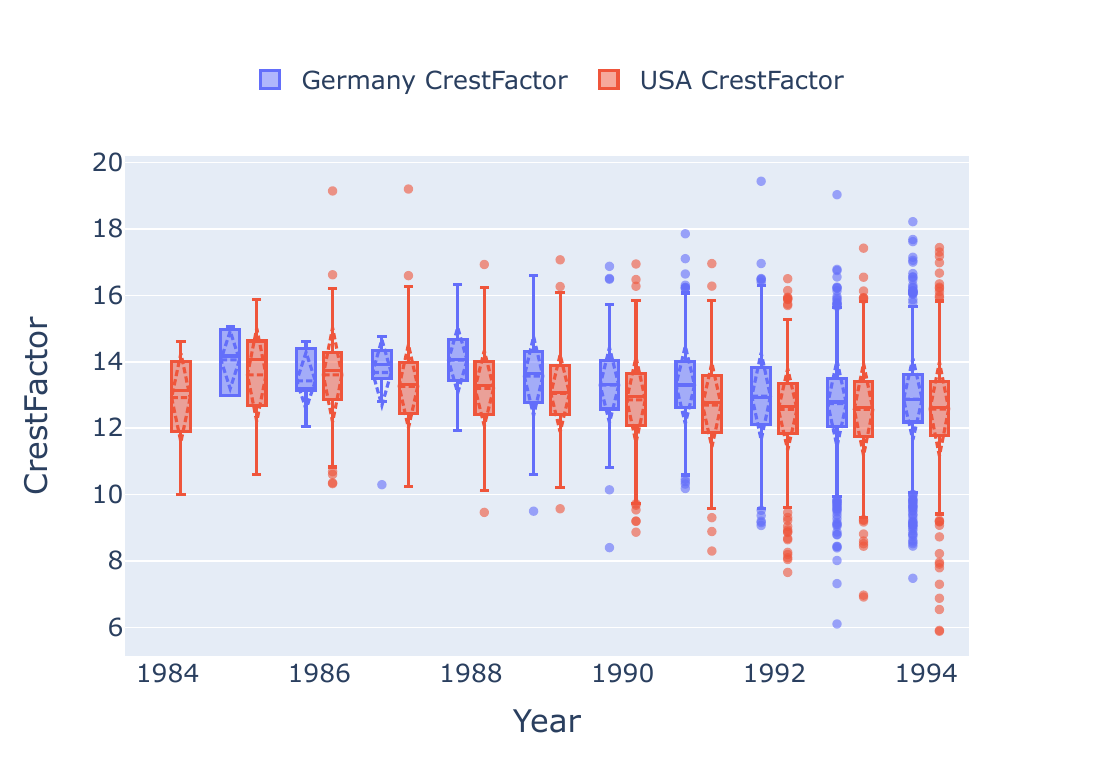}
\caption{Boxplots of median crest factor in Germany and America over the years.}
\label{pic:boxcrest}
\end{figure}

Figure \ref{pic:boxcrest} shows the median crest factors. In both countries, the crest factor gradually reduces, indicating a tendency towards (hyper-)compression or towards more melody and harmony compared to percussive sounds.

\begin{figure}[hbt!]
\centering
\includegraphics[width=.49\textwidth]{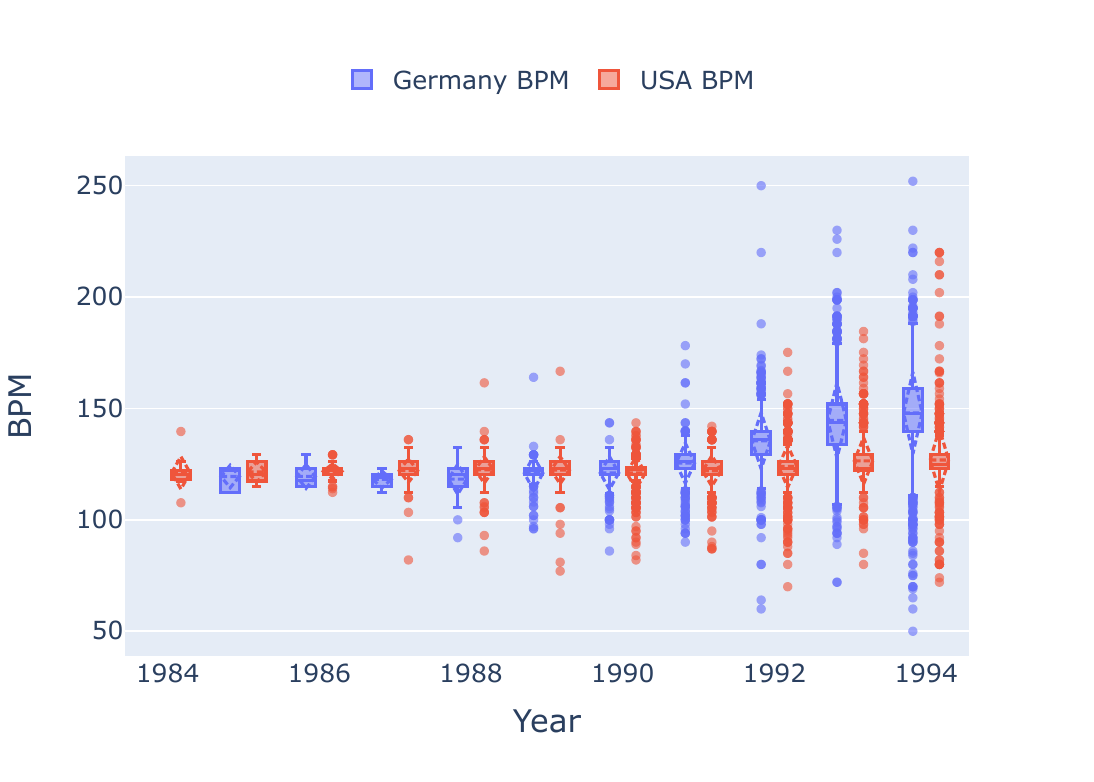}
\caption{Boxplots of bpm in Germany and America over the years.}
\label{pic:boxbpm}
\end{figure}

The boxplots of the bpm are shown in Fig. \ref{pic:boxbpm}. In Germany, the bpm rise over time. So does the spread. These two observations highlight that the German house and techno music evolved and diversified. In contrast, the tempo of American house and techno music is steady, and the spread is small. Only the magnitudes of outliers spread.

Despite significant differences identified via MANOVA, the boxplots do not provide a detailed insight into the differences between the nations, the years, and their combination. As a nonlinear dimensionality reduction and visualization method, a SOM may provide a clearer insight into the data and the nature of the differences between nations, years, and the different time evolutions between the nations.

\subsection{Self-Organizing Map}
\begin{figure}[hbt!]
\centering
\includegraphics[width=.49\textwidth]{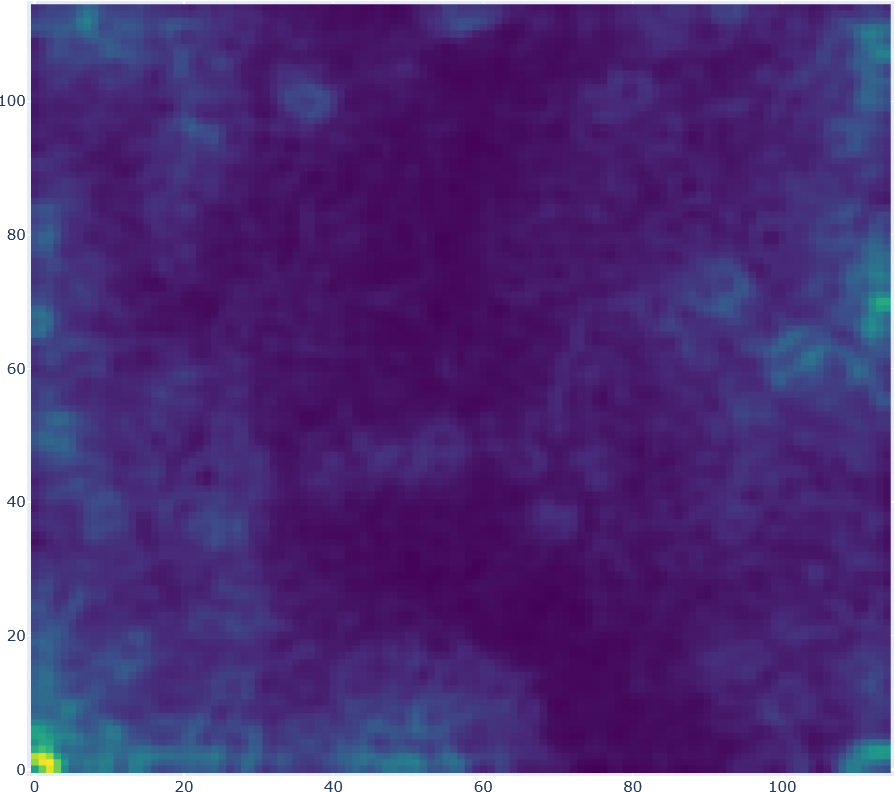}
\caption{Unit matrix (U-matrix) of the neural network trained with median bpm, phase space, channel correlation and crest factor of all 9,029 tracks.}
\label{pic:som}
\end{figure}

The U-matrix of the trained SOM is illustrated in Fig. \ref{pic:som}. On the left- and on the right-hand side, the SOM is turquoise, indicating that neighboring regions are more diverse than the dark-blue region in the middle that goes from top to bottom. There are only two separation lines. One clearly separates the unit in the lower-left corner from the rest, and one separates the units in the lower-right corner from the rest. This indicates that there are  mostly gradual transitions between the regions on the map, and not many sharp boundaries.

\begin{figure}[hbt!]
\centering
\includegraphics[width=.49\textwidth]{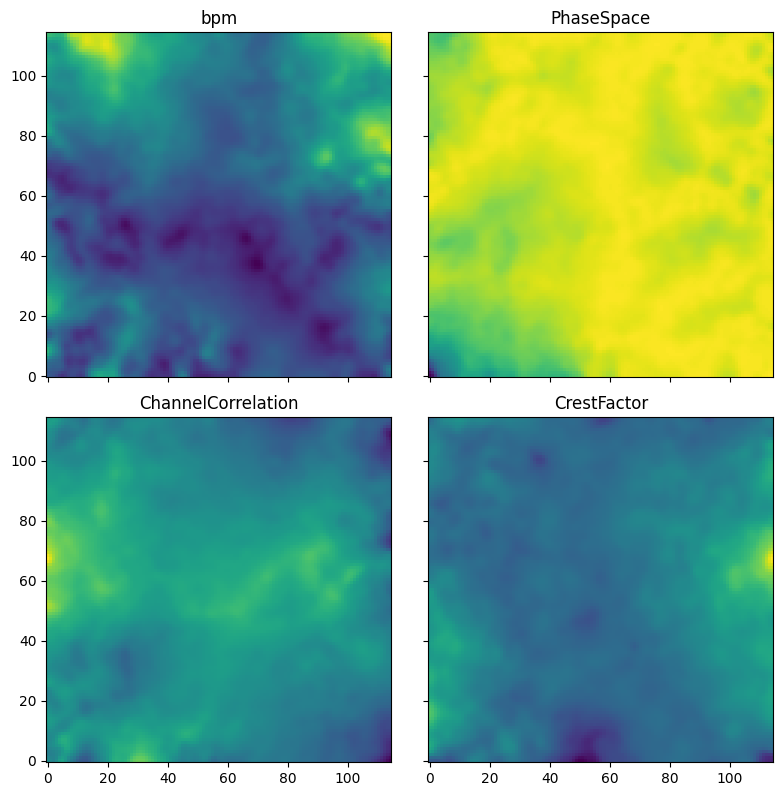}
\caption{Component planes, i.e., magnitude of the 4 components at each unit.}
\label{pic:components}
\end{figure}

What is similar and dissimilar in these regions can be seen in Fig. \ref{pic:components}, which shows the magnitudes of the $4$ component planes. Overall, the bpm varies a lot along the map, while the phase space has a large magnitude except for the lower-left corner. The channel correlation mostly has a medium magnitude, except for very low values in the upper and lower right, and very large values on the left, especially at medium height. The crest factor is mostly medium in the center. In the lower-middle, and some small regions, the crest factor is low. High values can be found on the left-, and on the right-hand side.

The dark blue region on the U-matrix contains music with a low to medium tempo (bpm), and a medium crest factor. A more detailed examination of the SOM will reveal the meaning of the regions. All SOMs illustrated in this paper are available as interactive maps under \url{https://timziemer.github.io/technoanalysis.html}.

\begin{figure*}[btp!]
\centering
\includegraphics[width=.99\textwidth]{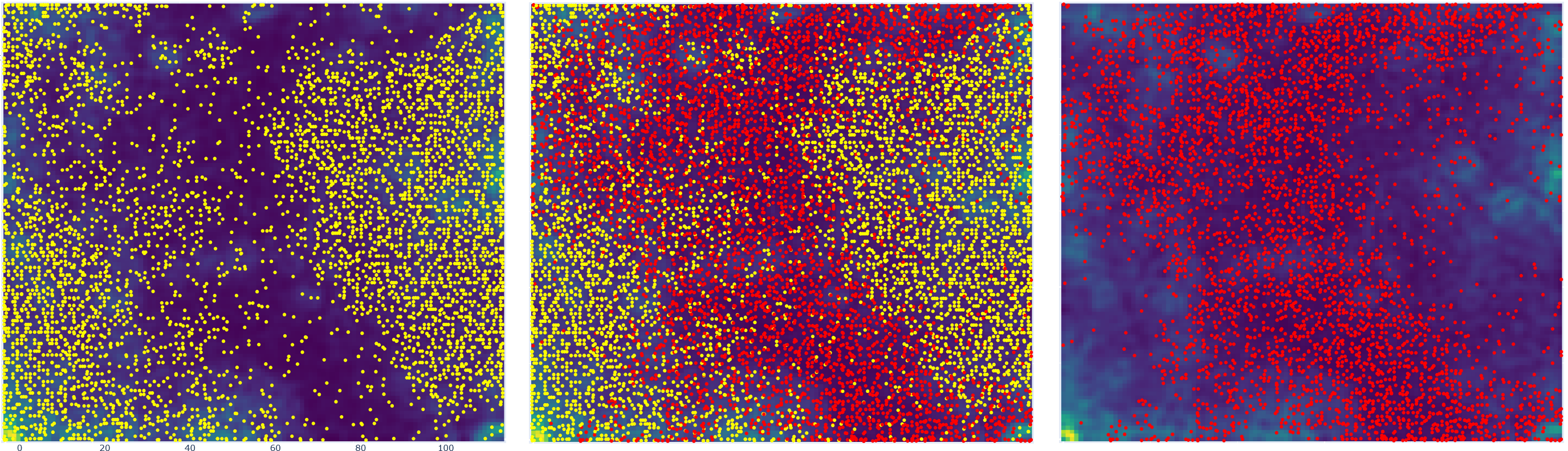}
\caption{Distributions of German tracks (left, yellow) US tracks (right, red) and both (center) on the SOM. The \href{https://timziemer.github.io/technomaps/2025-02-24-med-bpm-phsp-chacorr-crest-USvsGer.html}{interactive SOM} is available online.}
\label{pic:gbu}
\end{figure*}

In Fig. \ref{pic:gbu}, all tracks are assigned to their BMU. Overall, the German tracks and the US tracks are well-separated. Many US tracks are located along the dark-blue region, meaning that they are more alike than the German tracks in the turquoise regions. Naturally, tracks from both nations scatter into the region of the other nation. The segregate tracks in the lower-left corner are experimental pieces with a negative channel correlation and a low crest factor due to loud atmo sounds and pads like \emph{Wahnfried feat. Klaus Schulze - Abyss (G, 1994)}. The segregate tracks in the lower-right corner are ambient pieces without a beat, like \emph{Eradicator - Starving (G, 1994)}.

\begin{figure*}[btp!]
\centering
\includegraphics[width=.99\textwidth]{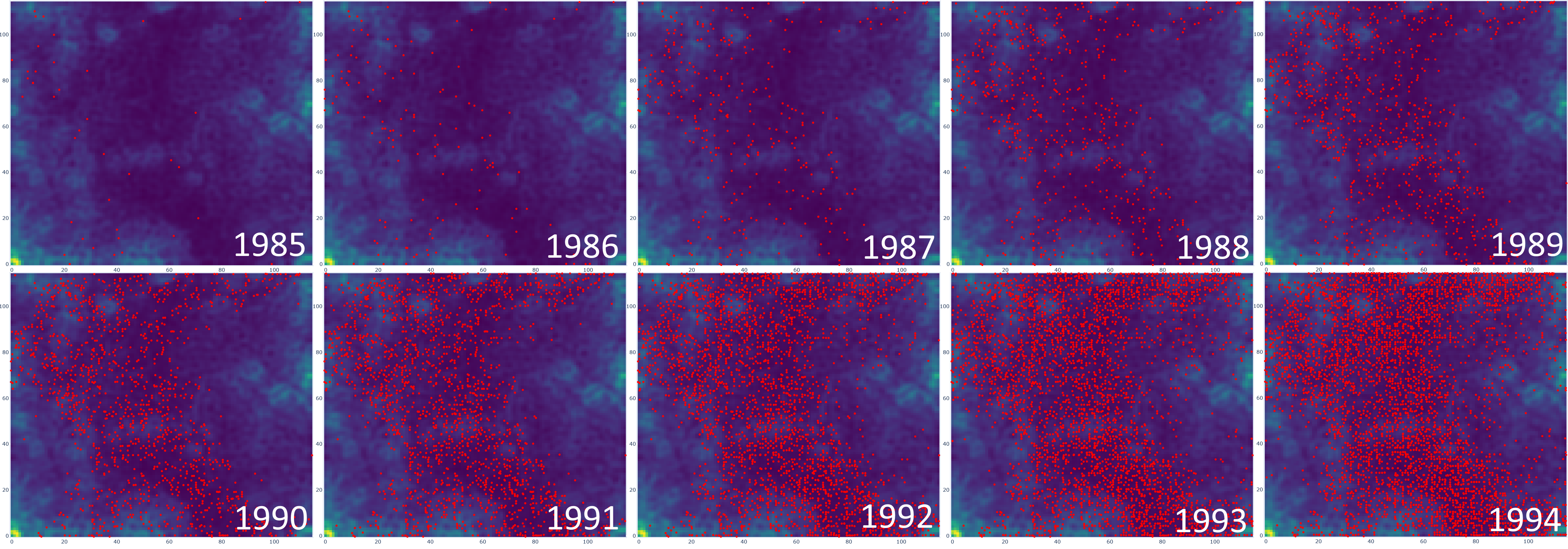}
\caption{US house and techno tracks over the years. Every year, the new tracks are added to the map. The \href{https://timziemer.github.io/technomaps/2025-02-24-med-bpm-phsp-chacorr-crest-Nation-Years.html}{interactive SOM} is available online.}
\label{pic:usayears}
\end{figure*}

The development of American house and techno music over the years is observable in Fig. \ref{pic:usayears}. By 1986, the typical US-region is already visible. In the following years, the region is filled more and more densely. This America-typical region has a medium to lo bpm value (tempo), a low crest factor, indicating dynamic range compression. Also, the channel correlation (indicating stereo spread) is quite alike). Only from 1992 to 1994, some outliers are observable. But the majority of tracks stays in the same region. What these outliers are, will become clear when observing the different styles of US house and techno music.

\begin{figure*}[btp!]
\centering
\includegraphics[width=.99\textwidth]{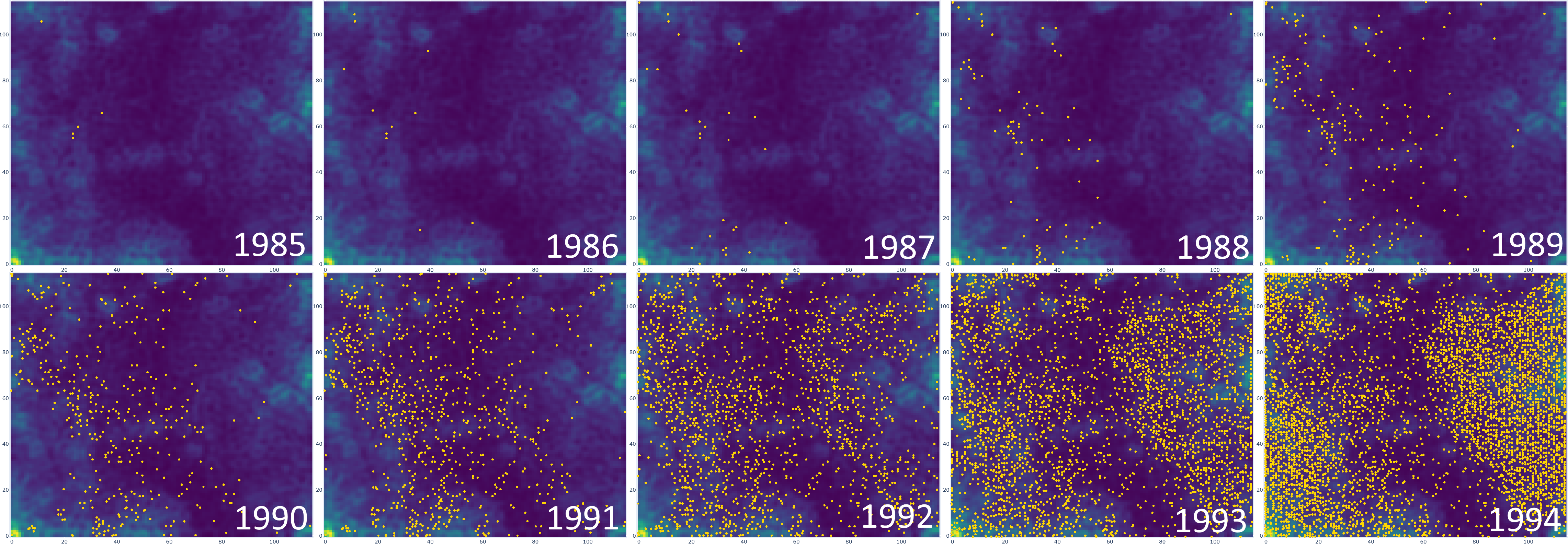}
\caption{German house and techno tracks over the years. Every year, the new tracks are added to the map. The \href{https://homepages.uni-paderborn.de/ruppert/technomaps/2025-02-24-med-bpm-phsp-chacorr-crest-Nation-Years.html}{interactive SOM} is available online.}
\label{pic:germanyears}
\end{figure*}

The evolution of German house and techno music is illustrated in Fig. \ref{pic:germanyears}. Very few points exist until 1988. Until 1990, the German tracks mostly fall in the same region on the SOM as the American tracks. In 1991, the region grows slightly. In 1992, suddenly, most tracks are allocated anywhere except the US region. This trend continues until 1994, where many tracks are located near the edges, far away from American tracks. This German-typical region exhibits a larger variety in terms of bpm, channel correlation, and crest factor.

\begin{figure}[hbt!]
\centering
\includegraphics[width=.45\textwidth]{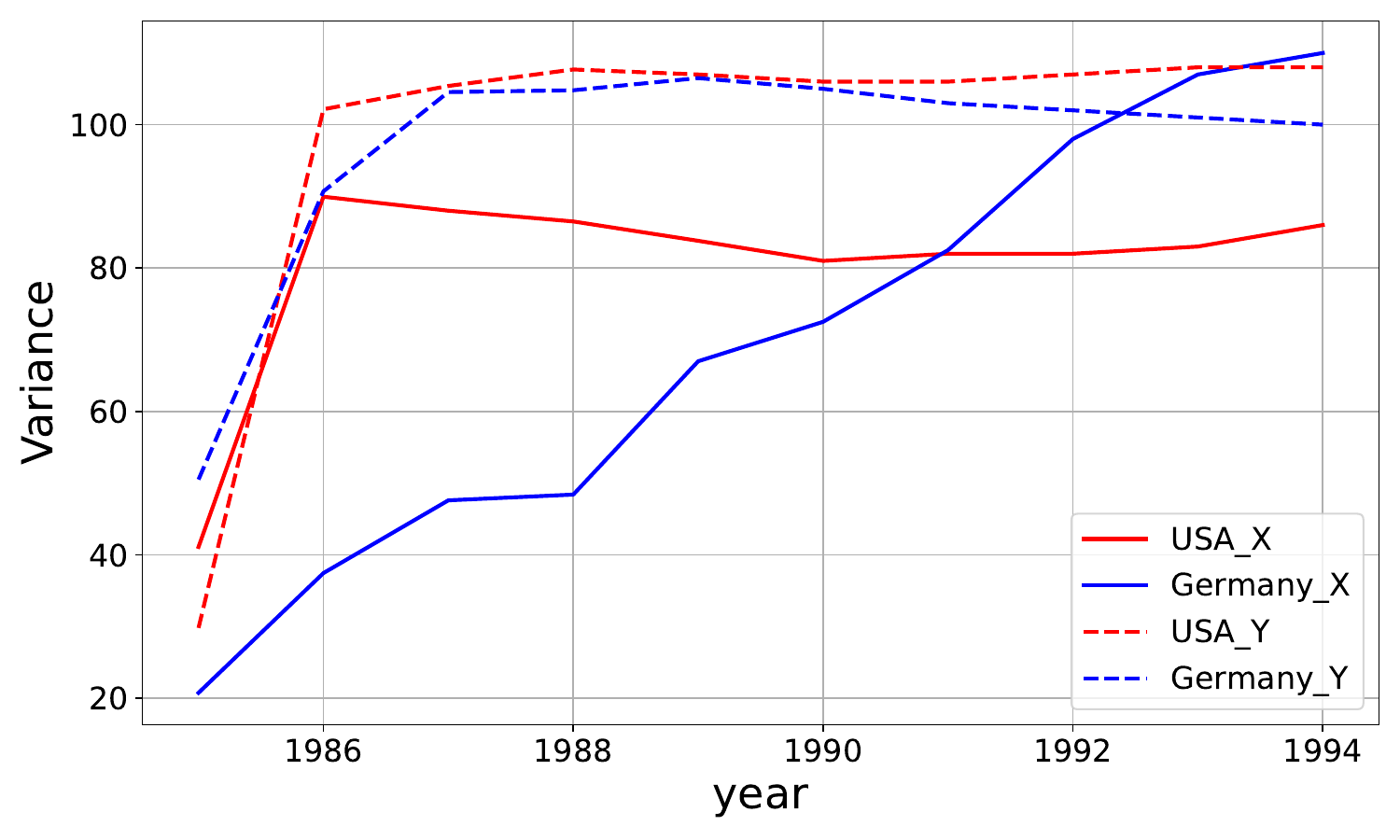}
\caption{Location variance of tracks on the SOM.}
\label{pic:var}
\end{figure}

These qualitative observations are quantifiable, too. Figure \ref{pic:var} shows that the $x$- and $y$-coordinates of American tracks stays constant over the years. In Germany, the variance of the $x$-coordinate rises continuously, underlining how newer tracks keep deviating more and more from the American tracks.

\begin{figure}[hbt!]
\centering
\includegraphics[width=.45\textwidth]{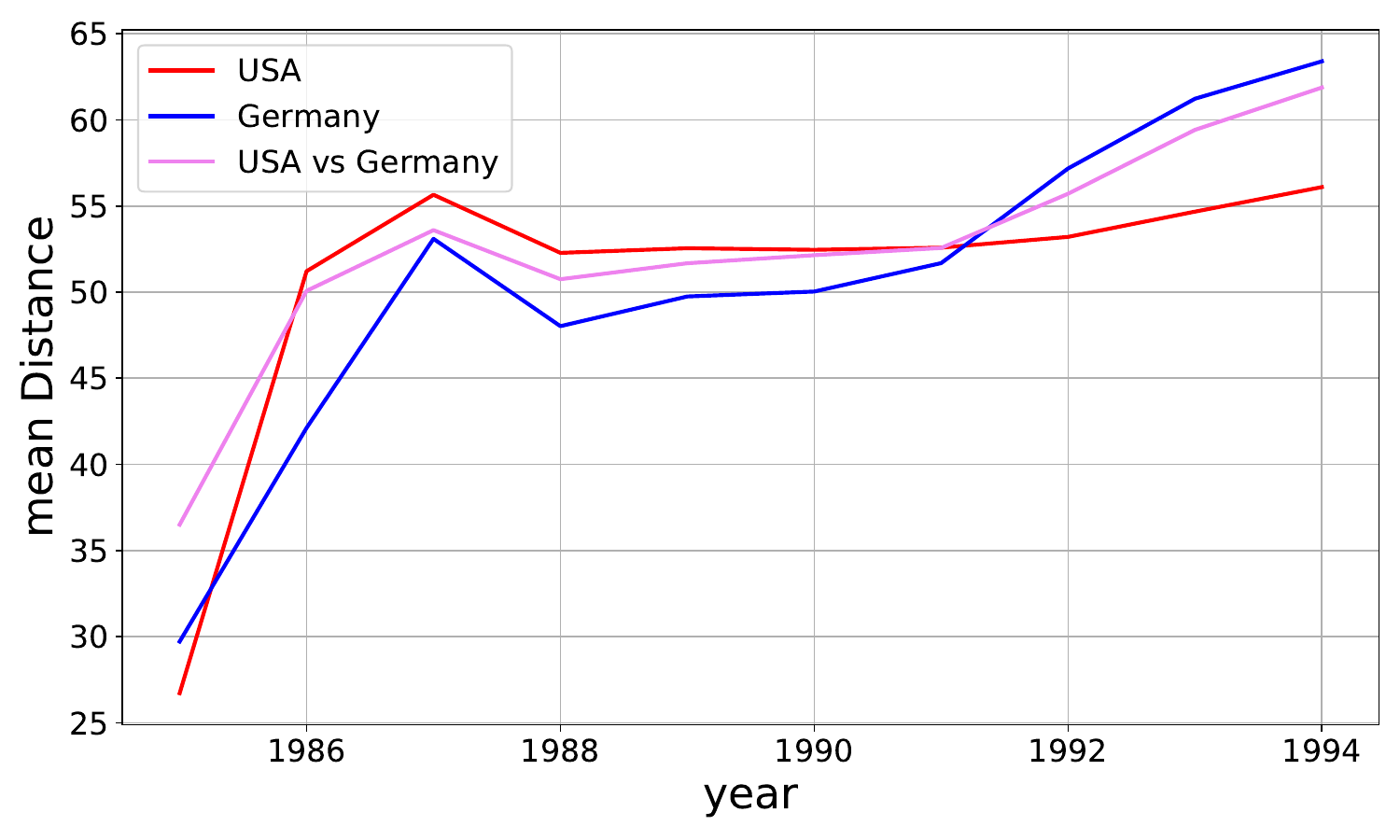}
\caption{Distances between tracks on the SOM.}
\label{pic:dist}
\end{figure}

The crucial year 1992 is also observable in the distances between the tracks plotted in Fig. \ref{pic:dist}. While the mean distance between all American tracks stays fairly constant over the years, the German tracks start growing apart between 1991 and 1992. And so do the distances between the German and American tracks. From 1992 to 1994, the German tracks are almost the opposite of the American tracks concerning their location on the SOM.

\begin{figure*}[hbt!]
\centering
 \includegraphics[width=.99\textwidth]{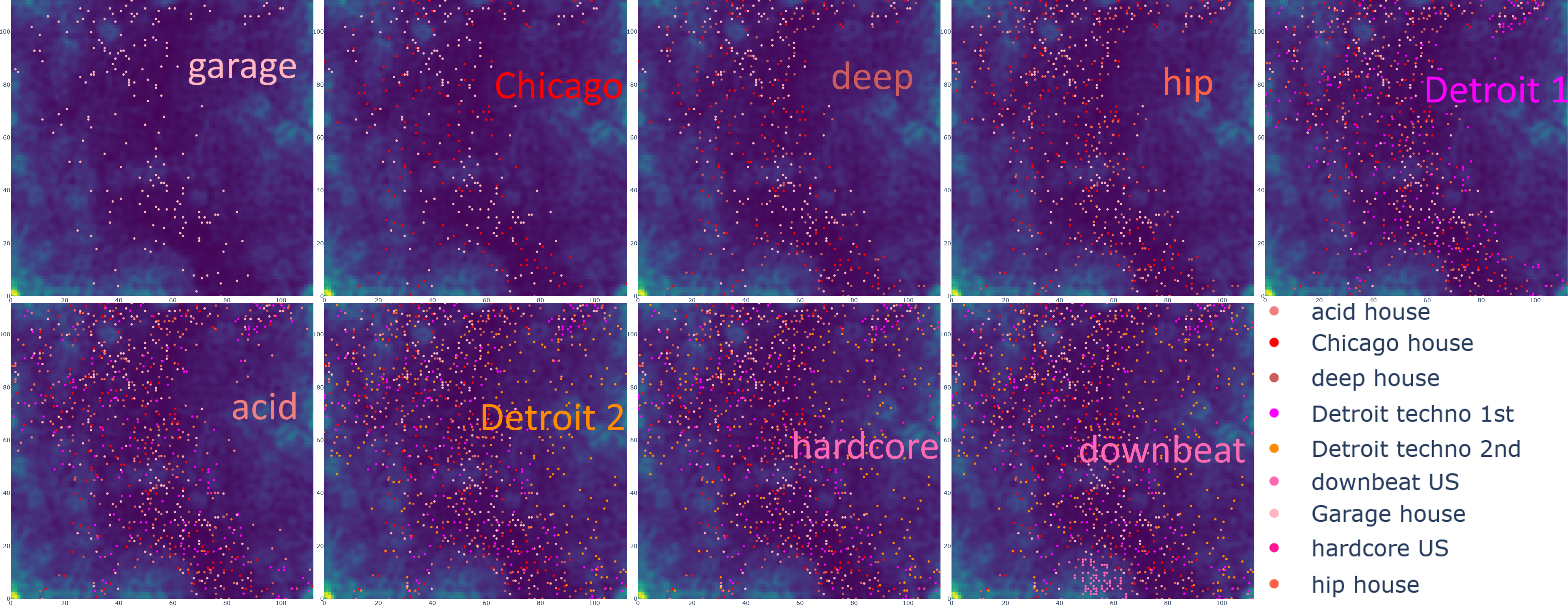}
\caption{9 different styles of US house and techno music. The \href{https://timziemer.github.io/technomaps/2025-02-24-med-bpm-phsp-chacorr-crest-Genres.html}{interactive SOM} is available online.}
\label{pic:usagenres}
\end{figure*}

Different styles of American house and techno music are illustrated in Fig. \ref{pic:usagenres}. The locations of  garage house, Chicago house, deep house, hip house and the first wave of Detroit techno are almost identical. Acid house largely overlaps with the rest, but spreads a little more. The second wave of Detroit techno spreads even more and has less overlap with the other styles. Hardcore (in terms of hardcore/gabba, not UK hardcore/drum \& bass) occupies a small region on the right-hand side (near the ``core'' and a bit above, where the bpm and the crest factor are the highest in Fig. \ref{pic:components}). And downbeat has its own island in the middle of the lower edge (where the crest factor and the bpm are the lowest, see Fig. \ref{pic:components}). Overall, most US styles are much alike regarding the recording studio features. Exceptions are the second wave of Detroit techno, hardcore, and downbeat.

\begin{figure*}[hbt!]
\centering
\includegraphics[width=.99\textwidth]{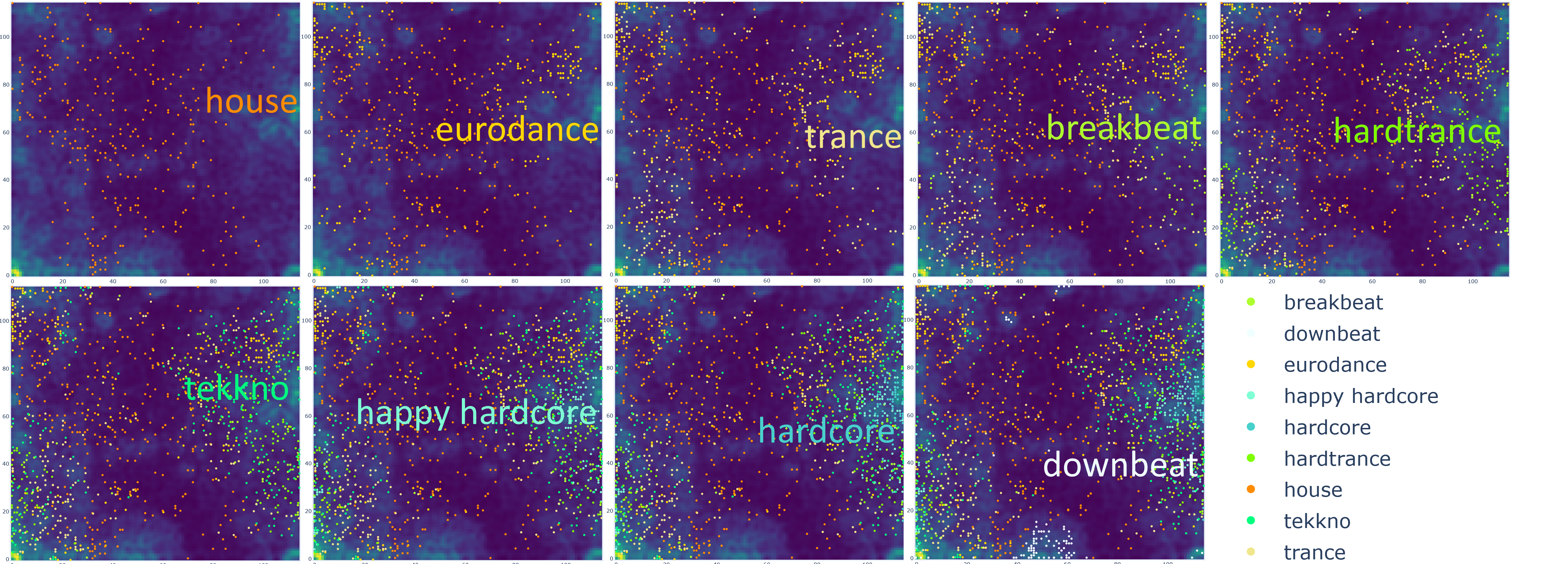}
\caption{9 different styles of German house and techno music. The \href{https://homepages.uni-paderborn.de/ruppert/technomaps/2025-02-24-med-bpm-phsp-chacorr-crest-Genres.html}{interactive SOM} is available online.}
\label{pic:germangenres}
\end{figure*}

Different styles of German house and techno music are illustrated in Fig. \ref{pic:germangenres}. The German house music spreads over the complete American region. Eurodance lies near the edges of house music and slightly beyond. Trance has some overlap with eurodance, but also occupies a new region in the lower-left. Breakbeat has a wide spread, typically outside the house region. The same is true for hardtrance, which often lies near trance, but further away from the house region, more towards the corners and edges. The same is true for tekkno. Happy hardcore fills the space between the hardtrance and tekkno spread. Hardcore is centered in the same region as US-hardcore, and downbeat lies on the US-downbeat island. Overall, German house, hardcore, and downbeat resemble the US counterpart. The other styles are partly similar to each other (e.g., hardtrance and tekkno), but mostly different from each other and from US house and techno music.

\subsection{Random Forest Classifier}
\begin{figure}[hbt!]
\centering
\includegraphics[width=.49\textwidth]{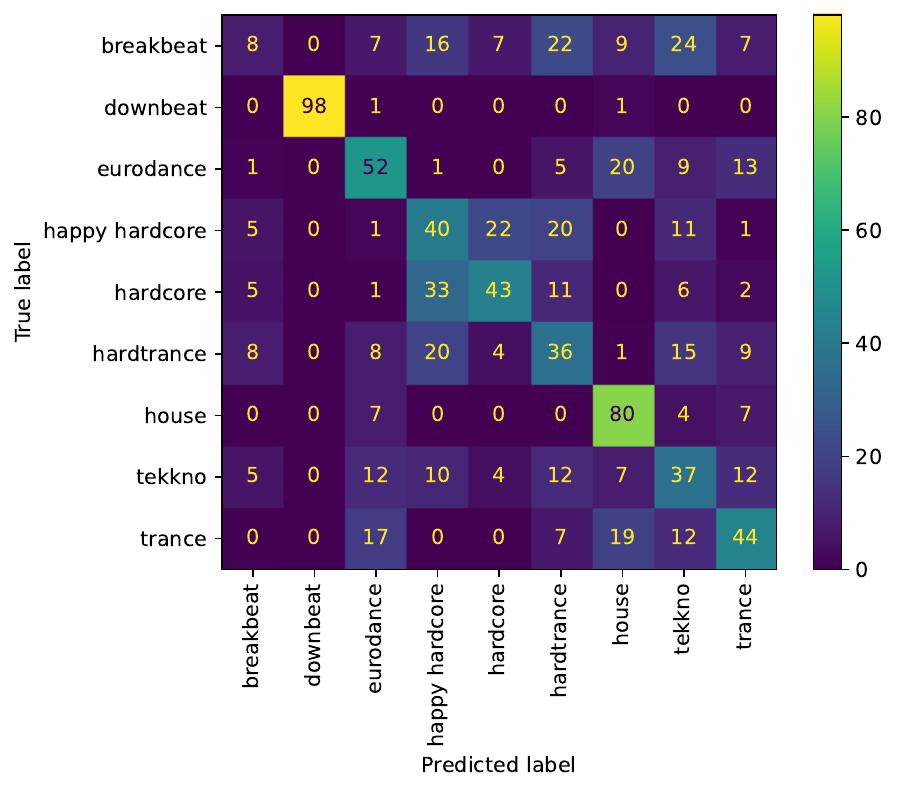}
\caption{Confusion matrix of German dance music styles according to a random forest classifier.}
\label{pic:rfgerm}
\end{figure}

The German styles are analyzed using a random forest classifier with 100-fold cross validation. The confusion matrix can be seen in Fig. \ref{pic:rfgerm}, summarized in percentages. The classifier had a training accuracy $= 0.983\pm0.003$, and a test accuracy $= 0.515 \pm 0.018$, precision $ = 0.503 \pm 0.017$, recall $ = 0.515 \pm 0.017$, and f1-score $ = 0.506 \pm 0.017$. The classifier can distinguish the styles fairly well. The diagonal is clearly visible, highlighting that breakbeat is the only style that is classified with a recall (sensitivity) below the chance level of $1/9\approx11$\%, while all other styles exhibit a recall between $36$ and $98$\%. Naturally, downbeat has a recall of $98$\%, because the bpm are distinctly lower. But also house music is recognized with a recall of $80$\%, confirming the distinct region on the SOM. Happy hardcore and hardcore are confused quite often, which is expected because the happiness certainly lies in the melodies rather than in the sound. Breakbeat is not recognized well, highlighting that the considered recording studio features describe the sound dimensions, not the rhythm.

\begin{figure}[hbt!]
\centering
\includegraphics[width=.49\textwidth]{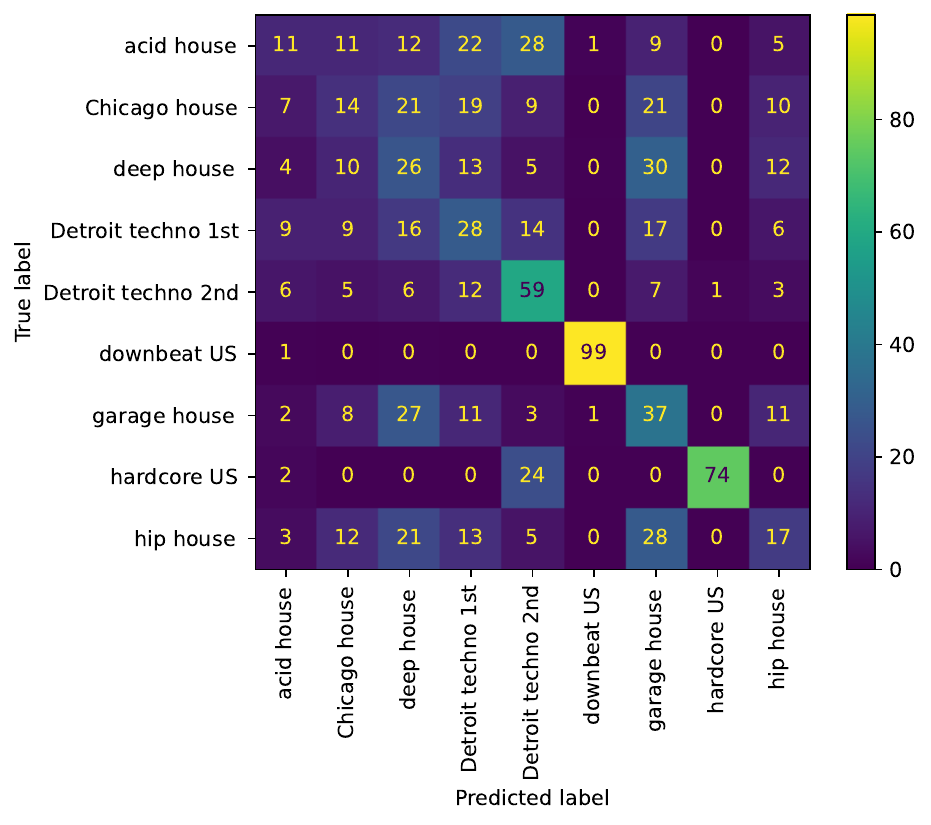}
\caption{Confusion matrix of American dance music styles according to a random forest classifier.}
\label{pic:rfusa}
\end{figure}

The American counterpart is illustrated in Fig. \ref{pic:rfusa}. The classifier had a training accuracy $= 0.991\pm0.002$, and a test accuracy $  = 0.369 \pm 0.020$, precision $ = 0.362 \pm 0.019$, recall $ = 0.369 \pm 0.020$, f1-score $ = 0.362 \pm 0.019$. Overall, the classifier cannot distinguish the styles well. The diagonal is less pronounced, and even though only acid house performs at chance level with a recall below of $11$\%, the other styles exhibit a recall between $14$ and $99$\%. As in Germany, downbeat has an exceptionally high recall of $99$\%, owed to the bpm feature. Hardcore is also recognized well, because -- in contrast to German hardtrance, happy hardcore, some eurodance and tekkno -- almost no other tracks have such a high bpm. As observed in the SOM, the second wave of Detroit techno is also recognized, as its sound left the typical house region. Garage house also distinguished well from the rest. Hip house, Chicago house, and deep house are confused quite often, and acid house is not recognized at all.

Overall, the random forest classifier confirms that the German house and techno styles are much more distinct than the US American styles concerning the recording studio features.

\section{Discussion}
The data analyses based on MANOVA, SOM and random forest highlight three things: 1.) House and techno tracks from Germany and the United States of America are quite different. According to the MANOVA, the differences are statistically significant, if we model the distributions of feature magnitudes as normal distributions. Even though the neural network was only fed with the median of recording studio features, each nation mostly occupies a different region on the SOM, with decreasing overlap, particularly from 1992 onward. No metadata about nation, year, style, etc. was fed to the neural network. This means that one can tell German and American house and techno tracks apart by recording studio features, i.e., aspects of music production and mixing. The boxplots and the component planes (see Fig. \ref{pic:components}) show that mostly the bpm behave very different between the countries, but also the other features exhibit differences, particularly the spread of ChannelCorrelation and CrestFactor.

2.) Many American house and techno tracks are very similar to each other according to the recording studio features. In the SOM (see Fig. \ref{pic:gbu}), US music is mostly concentrated on the dark blue region, indicating high similarity. Moreover, Fig. \ref{pic:usagenres} shows that garage, Chicago house, deep house, hip house and the first wave of Detroit techno are very much alike. Only some acid tracks are a bit different, just like tracks from the second wave of Detroit techno, hardcore and downbeat. \citep{collins} also observed that Detroit techno is more heterogeneous than Chicago house based on rather different features. But their classifiers distinguished them at chance level. Some styles sound similar in the US and Germany according to our features, like house (e.g., \emph{Farley 'Jackmaster' Funk - Love Can't Turn Around (Club Mix) (U, 1989)} and \emph{The Scandalous Tribe Feat. Vamps'n Roses Introducing Damon - Yes Sir, I Can Boogie (Full-Vocal-Speed-Mix) (G, 1990)}), hardcore (e.g., \emph{Dj Skinheads - Extreme Terror (The Gangster Mix) (U, 1994)} and \emph{E-Legal - Defy Hell (G, 1994)}), and downbeat (e.g., \emph{Jamie Principle - If it's love (U, 1992)} and \emph{Bombast Broz - Listen to My Music (Demo Version) (G, 1990)}).
 The other styles are distributed over turquoise regions, indicating that the tracks are less alike. While breakbeat and tekkno show much overlap with each other, the other styles have more distinct regions. The random forest classifier confirms this observation with $52$\% vs. $37$\%  classification accuracy.

3.) US house and techno does not change much over time regarding the recording studio features. Even though the spread in the boxplots increases between 1992 and 1994, the majority of tracks stays within the same magnitude region. Likewise, the American tracks stay in the same region on the SOM. In Germany, the situation is different: On the SOM, the German tracks are located in or near the US-region until 1990. In 1991, the region starts growing, i.e., German music is becoming more and more distinct. Between 1992 and 1994, most tracks lie outside the US region, more and more towards the corners and edges. In this period, the German tracks are located at the opposite region compared to the American tracks. Thus, 1992 may be considered a revolution as observed by \citep{mauch}. Figure \ref{pic:var} highlights how the German tracks vary more and more from each other each year, and Fig. \ref{pic:dist} shows that the distance among German tracks and between German and American tracks grows from 1991 on.

While MANOVA and boxplots reveal significant differences and show some distribution details, it is the SOM and the random forest classifier that provide a deep insight into the nature of the tracks and their differences over nations, year, and style.


The three observations indicate in what respect house and techno music developed differently in the USA and Germany. An important observation is that the dissipation from the original house path started in 1991, and the diversification of German house and techno music is visible in 1992, i.e., two years before the breakthrough. So this dissipation and diversification may be causes rather than results of the breakthrough.

The audio analysis supports many claims that protagonists made.
\subsection{Narratives vs. Features}
\label{verification}
As presented in Table \ref{tab:observations}, people from the early house and techno scenes made several statements. Our audio analysis supports (\cmark) or challenges (\xmark) some statements, but provides no further insight into others (\omark).

For example, \citep[p. 16]{volkweintechno} observed that early German house music emulated American house. 
In the SOM, most German tracks before 1991 overlap with the region that is typical for American tracks. In an interview in \citep[min. 37]{overtheworld}, the Chicago house pioneer 
Marshall Jefferson stated: ``We did extremely good for our talent level. But overall, I don’t think we had the musical talent to progress to that next level. Other people took it to the next level.'' And it seems that except for the second wave Detroit techno, and the small hardcore and downbeat scenes, it was mostly German music that left the established house music path and created a new sound.

According to \citep{technodeep}
, German DJs like DJ Hell and DJ Rok anticipated in 1992 that techno would experience new creativity and inspiration. Sven Väth, a veteran of the Frankfurt club scene, stated that the Germans found ``their'' techno in 1992, which became better and internationally recognized. 1992 is also considered the birth year of trance, Frankfurt the birthplace \citep{volkweintechno}. The MANOVA reveals that the analyzed German and US music is significantly different. Moreover, the SOM shows that many German track from 1992 to 1994 lie outside the American region. However, the boxplots show the widest distribution of phase space, channel correlation, and bpm in Germany from 1992 to 1994, indicating that \emph{the} German techno is actually quite divers. Likewise, the German tracks show a large spread between 1992 and 1994 in the SOM, highlighted also in Fig. \ref{pic:dist}.

\cite{sicko} stated that the German scene was more open and dynamic. On the SOM, German tracks show a transition from the central, original region towards new regions near the edges of the map. And the RF classifier can distinguish the newly arising German styles better from one another than the different American styles. These observations support the statement that the German house and techno music was more dynamic, or at least with more transition and distinction.

According to Wolle XPD, there was consensus that megaraves like Mayday would make all DJs play the same and all music sound the same \citep{wick}. Wolle XPD is a German DJ and organizer of the Tekknozid events. Tekknozid was an early megarave, where three DJs would play for thousands of ravers in a dedicated night. In contrast, Mayday has dozens of DJs. Mayday events started in 1991. Others agree that the masses cannot create something new \citep{berlin}. On our SOM, German music starts diverging from its origins in 1991 and diverges and diversifies in 1992 to 1994. The RF models can distinguish the German styles well. Many of these styles emerged in the early 1990s. These findings challenge the statement that megaraves made make all German house and techno music sound the same. While our data does not examine what DJs have played at megaraves, the music in general sounded more and more diverse, regarding the recording studio features.


According to \citep[chap. 12]{flash}, house and techno music in the US failed at establishing unconventional branches from the house and techno tradition, like jungle in England, gabba in the Netherlands, or trance in Germany. 
WestBam, one of the first German techno DJs and founder of the first German techno label, stated that the Detroit ``purists'' stayed among themselves, which is why Detroit never got a vivid techno scene compared to Berlin \citep{planetwissen}
. Likewise, DJ Pierre, a Chicago house DJ credited as the inventor of acid house, stated that the Chicago artists were competitors as DJs and producers, but felt like teammates \citep{overtheworld}
. In Germany, rivalry between Frankfurt and Berlin
, and the dissociation of different styles 
fostered the establishment of Germany's scenes and styles \citep{familie,loveparade}. 
A lot of this rivalry will have taken place outside the music, expressing itself in fanzines, clubs owners and event organizers booking particular DJs, record labels signing particular artists, DJs playing and avoiding particular music, and record stores pre-selecting the music to sell to DJs (c.f.\citep{szene,volkweintechno,Kaul2017}). Still, the MANOVA and boxplots, SOM, and RF confirm that German styles started similar to the US sound, but became more diverse in terms of \emph{bpm}, \emph{PhaseSpace}, \emph{ChannelCorrelation} and \emph{CrestFactor}, while US styles hardly left the traces of traditional house. Moreover, the RF classifier cannot distinguish most American styles. 

The fact that the second wave of Detroit techno also shows quite a spread, away from the old house music sound, is not a contradiction to WestBams observation. Mike Banks, co-founder of Underground Resistance and the second wave of Detroit techno, called Detroit techno an export product that was particularly famous in Berlin, cf. \citep{berlintechno}
. And \citep{Kaul2017} 
confirms that Underground Resistance oriented themselves to the European market. Jeff Mills, another second wave Detroit techno DJ, also had a larger audience in Berlin than in the United States \citep{szene}
. He stated that late 1980s Detroit was inspired by Belgian EBM music, explaining why Underground Resistance sounded so similar to some European acts \citep{flash}. In the SOM, we can see that the second generation of Detroit techno spreads strongly into the German region.

One common narrative is not reflected in the SOM: the German reunification, and particularly the fall of the Berlin Wall in late 1989, was often mentioned as a key moment in the German techno scene (e.g. \citep[pp. 72-7]{technodeep,szene,loveparade}). 
Protagonists emphasize how the scene grew fast, and how ravers from East Berlin brought a new spirit, and how long-abandoned industrial halls in the restricted border zone were used as nightclubs and for ever-growing raves. However, the analyzed features do not show dramatic changes in German house and techno music in 1989 or 1990. One explanation may be that this historic event affected the scene, organization, and consumption of the music rather than the music itself. Another explanation is that it simply took time to take effect; maybe until 1992, where the SOM shows a dramatic change. But it is also likely, that this important event affected aspects of the music that are not reflected in the median values of the bpm, phase space, channel correlation and crest factor. More in-depth data examination should be dedicated to identifying how the German reunification affected the music.


In summary, analyzing over $9,000$ early house and techno tracks from Germany and the United States of America confirmed that Germany developed its own style in 1992, which was diverse and evolving, while American music sounded more alike and stagnated. This observation is based on recording studio features that represent sound aspects of the music production, not the structure of the tracks, the melody, harmony, or lyrics. While the sound is certainly only one aspect out of many that paved the way for German techno as a mass phenomenon and a flourishing scene, this study supports or challenges protagonists' statements based on audio analysis.

\subsection{Limitations}
The strengths of the recording studio features are that a) they represent sound aspects that matter to music producers, b) are utilized by music producers through audio monitoring tools during the creative process of music making, c) their magnitudes are directly interpretable sound parameters, and d) they are mutually independent. Still, they are incomplete sound descriptors. For example, they represent little information about timbre. One prominent result is that acid house has not clustered in our SOM, and has not been recalled by the classifier. This is certainly owed to the fact that the main identity characteristic of acid house is the combined use of a cutoff and a heavy resonance filter on a bassline, often realized through the Roland TB-303 synthesizer. The features fail at identifying this sound aspect.
%

Sound is not the only musical aspect that matters. While \citep{hawkins} states that rhythmic patterns are simple, \citep{honingh} highlights the importance of rhythm in dance music. Our set of recording studio features hardly represents aspects of rhythm. Consequently, breakbeat does not cluster in the SOM (Fig. \ref{pic:germangenres}) and the random forest classifier (Fig. \ref{pic:rfgerm}) cannot recall breakbeat. Adding a meaningful and interpretable rhythm feature may improve the audio representation and provide further insights.

And while audio features are often considered objective, the selection of features has a subjective component.
Further house and techno studies are required to validate whether the trend from imitation towards diversification is unique to Germany, or observable in other nations with house/techno affinity, too. And further studies on other genres are required to validate whether this trend was unique to house/techno or whether it applies to other music fashions, too. 
Of course, analyzing the music does not reveal how musicians influence each other, for example musicians moving from Detroit to Berlin, releasing tracks on German labels and importing and exporting records \citep{flash}.

Naturally, the audio aspect is just one part of the house and techno scenes or sub-cultures. The infrastructure in terms of record labels, nightclubs, radio stations, record and fashion stores, fanzines, advertisements, etc. are additional factors that are not considered in the present study. And of course, a lot of music may have never been released, many releases have hardly been played on raves, and a lot of music from various countries has been played in America and Germany, too. Moreover, the political dimensions, such as gay liberation, drug consumption, Afrofuturism, and civil rights movement \citep{society}, are not considered in the present study. 


\section{Conclusion}
Over $9,000$ early house and techno tracks from Germany and the United States have been analyzed using recording studio features and inferential statistics, a Self-Organizing Map (SOM), and a random forest classifier. The analysis revealed differences between the music of the two nations, various styles, and their development over time. While early German house and techno music was similar to the American music, it started diversifying in 1991 and segregating in 1992. The German house and techno styles are much more diverse than the US styles concerning recording studio features, i.e., aspects of music production and mixing. This diversification and segregation preceded the breakthrough of German techno, indicating that it was a catalyst, rather than a result of the breakthrough.

These differences, observed through content-based music analysis, are largely in accordance with statements of protagonists of the scenes and add a sound-based perspective on why the scenes developed so differently between the two nations. The connection between the statistical analyses and the interview statements show how the music reception and scene development relate to retrievable music production and mixing aspects. Methods from the field of music information retrieval, like big data analysis using feature extraction and machine learning, provide an audio-based enhancement of conventional music research for both exploratory and quantitative studies. The fact that the examination of big data audio analysis can yield similar results to interviewing protagonists indicates that audio analysis is a valid and insightful method to study the development of music scenes.


The approach can be transferred to other nations like the United Kingdom and the Netherlands, scenes, like breakdance, and arising or declining trends, like dubstep or American EDM. 
If the breakthrough of techno in Germany emerged because the music evolved away from its origins and diversified, a practical application could be to predict whether trending music will establish or decay, which helps record labels and event organizers plan accordingly. However, further studies are necessary to confirm or challenge the hypothesis that the evolution and diversification are an indicator of a breakthrough.

\section*{Data Availability Statement}
The HOTGAME house and techno music corpus, the feature extraction algorithms, and the data analysis codes are available online for further studies on \url{https://github.com/ifsm/HOTGAME-feature-extraction} and \url{https://github.com/ifsm/Techno-Analysis}.
\bibliography{sample}
\end{document}